\documentclass[usenatbib,useAMS]{mn2e}
\usepackage[T1]{fontenc}
\usepackage{aecompl}
\usepackage{graphicx}
\usepackage{amssymb}
\usepackage{longtable}
\usepackage{lscape}
\title[On the non-detection of $\gamma$-ray MSPs]{On the non-detection of $\gamma$-rays from energetic millisecond pulsars -- dependence on viewing geometry}

\author[Guillemot \& Tauris]{
  L. Guillemot,$^1$$^,$$^2$\thanks{E-mail: lucas.guillemot@cnrs-orleans.fr} and T.~M.~Tauris,$^3$$^,$$^2$ 
  \\
  $^{1}$Laboratoire de Physique et Chimie de l'Environnement et de l'Espace -- Universit\'{e} d'Orl\'{e}ans / CNRS, F-45071 Orl\'{e}ans Cedex 02, France
  \\
  $^{2}$Max-Planck-Institut f\"ur Radioastronomie, Auf dem H\"ugel 69, D-53121 Bonn, Germany
  \\
  $^{3}$Argelander-Insitut f\"ur Astronomie, Universit\"at Bonn, Auf dem H\"ugel 71, D-53121 Bonn, Germany 
  \\
  }

\begin{document}

\maketitle

\begin{abstract} 

Millisecond pulsars (MSPs) and normal non-recycled pulsars are both detected in $\gamma$-rays. However, it appears that a much larger fraction of known energetic and nearby MSPs are detected in $\gamma$-rays, in comparison with normal pulsars, thereby making undetected $\gamma$-ray MSPs exceptions. In this paper, we demonstrate that the viewing angles (i.e. between the pulsar spin axis and the line of sight) are well described by the orbital inclination angles which, for binary MSPs with helium white dwarf companions,  can be determined using the relationship between the orbital period and the white dwarf mass. We use the predicted viewing angles, in complement with values obtained from other constraints when available, to identify the causes of non-detection of energetic and nearby MSPs from the point of view of beaming geometry and orientation. We find evidence for slightly different viewing angle distributions, and postulate that energetic and nearby MSPs are mainly undetected in $\gamma$-rays simply because they are seen under unfavourable (i.e. small) viewing angles. We finally discuss the magnetic fields of $\gamma$-ray detected pulsars and show that pulsars which are efficient at converting their rotational energy into $\gamma$-ray emission may have overestimated dipolar magnetic field strengths. 

\end{abstract}



\section{Introduction}
\label{sec:Intro}

In its first five years of activity, the Large Area Telescope (LAT), main instrument of the \emph{Fermi Gamma-ray Space Telescope} \citep{FermiLAT}, has detected pulsed $\gamma$-ray emission from more than 130 pulsars\footnote{A list of $\gamma$-ray pulsar detections is available at https://confluence.slac.stanford.edu/display/GLAMCOG/\\ Public+List+of+LAT-Detected+Gamma-Ray+Pulsars.}, revolutionizing our understanding of high-energy emission from pulsars. More than a third of these $\gamma$-ray pulsars are millisecond pulsars (MSPs), i.e. neutron stars with short rotational periods ($P \lesssim 30$ ms), thought to have been spun~up by the transfer of angular momentum via accretion of matter from a binary companion \citep{acrs82,bv91,tv06}. $\gamma$-ray MSPs are detected either by folding the $\gamma$-ray photon arrival times using ephemerides from radio timing measurements \citep[e.g.][]{Fermi8MSPs,Guillemot2012a,Espinoza2013} or discovered through radio searches of unassociated $\gamma$-ray sources, such as those listed in the \emph{Fermi}-LAT Second Source Catalog \citep{Fermi2FGL}, and for which $\gamma$-ray pulsations were later revealed by folding the LAT data using radio ephemerides of such newly discovered pulsars \citep[e.g.][]{rrc+11,gfc+12,Barr2013}. A first direct discovery of an MSP in the \emph{Fermi} LAT data has also been reported \citep{Pletsch2012}. MSPs thus represent an important subpopulation among $\gamma$-ray pulsars, the dominant class of GeV $\gamma$-ray sources in our Galaxy \citep[see the Second \emph{Fermi} Large Area Catalog of $\gamma$-ray pulsars;][]{Fermi2PC}.

As is also the case for the normal population of pulsars, the detected $\gamma$-ray MSPs tend to be nearby and energetic objects, with large values of the spin-down luminosity $\dot E = - I \Omega \dot \Omega = 4 \upi^2 I \dot P / P^3$, where $I$ denotes the moment of inertia, generally assumed to be $10^{45}$~g~cm$^2$, $\Omega = 2 \upi / P$ is the angular velocity, and $\dot P$ is the spin-down rate. However, one important difference between the two pulsar populations resides in the much larger fraction of MSPs with high values of $\dot{E}/d^2$ that are detected in $\gamma$-rays, compared to the normal pulsar population. This large fraction of nearby and energetic radio MSPs detected in $\gamma$-rays, and also the lack of radio-quiet $\gamma$-ray MSPs, are interpreted as being due to the radio and $\gamma$-ray beams covering large and comparable fractions of the sky, in contrast to normal pulsars which have narrow radio beams \citep{Fermi2PC}. Radio and $\gamma$-ray detectabilities of pulsars and their dependence on beaming and $\dot E$ have also been discussed in e.g. \citet{Ravi2010}, \citet{Watters2011}, and \citet{Takata2011}.

Nevertheless, some high $\dot E / d^2$ MSPs escape detection in $\gamma$-rays, despite deep pulsation searches in the LAT data using highly precise radio ephemerides. For these undetected MSPs the distance may be larger than estimated; this can be the case for distances derived from models of the column density of free electrons in the Galaxy such as the NE2001 model \citep{NE2001,Deller2009}. They may also have unfavourable orientations, preventing the $\gamma$-ray beams from crossing our line of sight.

In this paper, we discuss the latter possibility by studying and comparing the viewing geometry of two samples of radio MSPs that are either detected or undetected in $\gamma$-rays, respectively. Pulsar geometry angles (namely the magnetic inclination angle between the spin axis and the magnetic dipole axis, $\alpha$, and the viewing angle between the spin axis and the line of sight, $\zeta$) are usually extracted from fits of radio polarization with the rotating vector model \citep[RVM;][]{RVM}, or from joint fits of radio and $\gamma$-ray pulse profiles in the context of geometrical models of emission from pulsars \citep[e.g.][]{Venter2012,Petri2011}. The former method generally does not work for MSPs, which exhibit complex polarization position angle variations that cannot be modelled with the RVM \citep[e.g.][]{Yan2011,Keith2012}, while the latter technique is only applicable to pulsars detected in both radio and $\gamma$-rays.

Here, we estimate viewing angles $\zeta$ of radio MSPs with helium white dwarf (He~WD) companions based on binary evolution arguments. Including this novel method strongly increases the total number of pulsars with estimated $\zeta$~angles and thus allows for a robust statistical analysis of $\gamma$-ray emitting MSPs in view of the $\zeta$ values. In Section~2, we present our method for estimating the $\zeta$ angle for MSPs in binary orbits with He~WD companions and test the validity of this method. In Section~3, we present our main results and investigate potential differences in the viewing angles of MSPs detected or undetected in $\gamma$-rays. Furthermore, we highlight a few individual MSPs and place interesting constraints on their masses or distances. Finally, we analyse and comment on the inferred dipole magnetic fields (B-fields) of efficient $\gamma$-ray emitting pulsars in Section~4. The latter sections are followed by a discussion of our results, and we summarize our findings in Section~5. 


\section{Estimating the viewing angles}
\label{sec: view}

\subsection{Predictions based on binary evolution}
\label{subsec: TS99}

It is well established that MSPs originate from low-mass X-ray binaries \citep[LMXBs;][]{acrs82,bv91,wv98,asr+09}. In LMXBs with initial orbital periods larger than a few days, the donor star will not fill its Roche~lobe until it is in the Hertzsprung gap or has moved up the red giant branch (RGB). For low-mass stars ($<2.3\,M_{\sun}$) on the RGB, there is a well-known relationship between the mass of the degenerate helium core and the radius of the giant star -- almost entirely independent of the mass present in the hydrogen-rich envelope \citep{rw71,wrs83}. This relationship is very important for the formation of binary MSPs because it results in a unique relationship between their final orbital period ($P_{\rm orb}$) and white dwarf mass ($M_{\rm WD}$) following the mass-transfer phase \citep{sav87,jrl87,rpj+95,ts99,DeVito2010,Shao2012}. The companions of these MSPs are He~WDs with masses $0.13 < M_{\rm WD}/M_{\sun} < 0.46$. The predicted correlation between $M_{\rm WD}$ and $P_{\rm orb}$ has previously been somewhat difficult to verify observationally since few MSPs had accurately measured masses of their companion star. However, over the past decade the correlation has been confirmed from mass measurements obtained from e.g. pulsar timing (Shapiro delay) or optical observations of He~WD companions \citep[e.g.][]{vbjj05}. This verification motivates us to apply the method presented in this study. As a consequence of loss of orbital angular momentum due to magnetic braking \citep[e.g.][]{vvp05}, LMXB systems with initial $P_{\rm orb}\lesssim 2\;{\rm d}$ are expected to be dragged towards each other and end up as close-binary MSPs with $P_{\rm orb}$ as short as a few hours \citep{esa98,prp02,db03}. Therefore, due to the still unknown strength of magnetic braking, the $M_{\rm WD}$ -- $P_{\rm orb}$~relation is less trustworthy for binary pulsars with $P_{\rm orb}<1\;{\rm d}$ (where He~WDs have masses between $0.13$ -- $0.18\;M_{\sun}$; the lower value is related to the Chandrasekhar--Sch\"onberg~limit), albeit still in accordance with observations. Some LMXB donor stars remain hydrogen rich and bloated which prevents them from terminating their mass-transfer process and forming a detached He~WD. These donors, which often suffer from ablation via the pulsar wind, can have their masses reduced significantly, leading to black-widow-type MSP systems \citep{rob13,ccth13}, or even complete evaporation and formation of an isolated MSP, in some cases possibly surrounded by an asteroid belt \citep{scm+13}.

The long time-scale ($10^8-10^9\;{\rm yr}$) of mass transfer in an LMXB is expected to cause the spin axis of the MSP to align with the orbital angular momentum vector of the system \citep{Hills1983,bv91}. Therefore, for such recycled pulsars the viewing angle is assumed to be equivalent to the orbital inclination angle, i.e. $\zeta = i$. Observations of the Doppler-shifted pulse signal yield the so-called mass function, $f$, of a binary pulsar obtained from $P_{\rm orb}$ and the projected semimajor axis of the pulsar orbit, $a_\mathrm{p} \sin i$. This mass function provides a relation between $M_{\rm WD}$, the MSP mass ($M_{\rm NS}$) and the orbital inclination angle of the system, $i = \zeta$:

\begin{eqnarray}
f(M_\mathrm{NS},M_\mathrm{WD},i) = \frac{4 \upi^2}{G} \frac{\left(a_\mathrm{p} \sin i\right)^3}{P_\mathrm{orb}^2} = \frac{\left(M_\mathrm{WD} \sin i\right)^3}{\left(M_\mathrm{NS} + M_\mathrm{WD}\right)^2}
\end{eqnarray}

\begin{figure}
\centering
\includegraphics[angle=-90,width=0.47\textwidth]{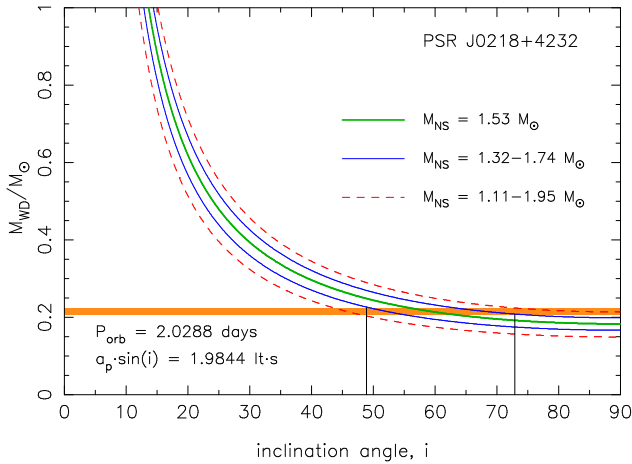}
\caption{For any given MSP orbiting an He~WD, and with measured mass function, $f(M_{\rm WD},M_{\rm NS},i)$, we can constrain the orbital inclination angle, $i$, of the system using the $M_{\rm WD}$--$P_{\rm orb}$ relation of TS99 for an assumed range of reasonable neutron star masses. Here is plotted the solution for PSR~J0218+4232. The measured $P_{\rm orb} = 2.03$ d yields the narrow interval of possible WD~masses ($0.207$--$0.225\;M_{\sun}$), as indicated by the orange horizontal band, from which we obtain $48\fdg8 < i < 72\fdg7$, assuming $M_{\rm NS}=1.32$--$1.74\;M_{\sun}$ (blue lines outlining the $1\,\sigma$ interval centred on $\langle M_{\rm NS} \rangle =1.53\;M_{\sun}$). See the text for further discussions.}
\label{fig:J0218}
\end{figure}

Here we use the $M_{\rm WD}-P_{\rm orb}$~relation of \citet{ts99}, hereafter TS99, to estimate $M_{\rm WD}$ from the observed $P_{\rm orb}$ of binary MSPs in the Galactic disc\footnote{We do not include MSPs in globular clusters since these binaries are embedded in a dense stellar environment and thus suffer from frequent exchange collisions and other encounters perturbing their orbits \citep{Heggie1975}.} and thereafter obtaining $\zeta$ values, using either measured constraints on $M_{\rm NS}$ or an assumed range of most probable values for $M_{\rm NS}$ (see Fig.~\ref{fig:J0218}). The results from this method can be tested directly against the $\zeta$ values obtained from the modelling of the radio and $\gamma$-ray profiles, as described below. As we shall now demonstrate, these two completely independent methods yield comparable results, which is not only another validation of the $M_{\rm WD}$--$P_{\rm orb}$~relation but also justifies calculating $\zeta$ values of MSPs for which no $\gamma$-ray pulsations are observed. In addition, it gives support to the hypothesis of MSP spin axes being aligned with their orbital angular momentum vectors.

\subsection{Viewing angle predictions for a sample of MSPs}
\label{subsec: zeta_calc}

To demonstrate that the TS99~relation provides a good estimator of pulsar viewing angles, $\zeta$, and to study the influence of $\zeta$ in the detectability of radio MSPs in $\gamma$-rays, we started by selecting Galactic disc MSPs (here defined as pulsars with $P < 30$~ms) from the 1.47 version of the Australia Telescope National Facility (ATNF) pulsar catalogue\footnote{http://www.atnf.csiro.au/people/pulsar/psrcat/} \citep{mhth05}. We selected MSPs in binary systems and likely to be orbiting He~WD companions, using the `BinComp' parameter, based on the criteria defined in the appendix of \citet{tlk12}. For each of the selected MSPs, spin-down rates $\dot P$, distances\footnote{Using the `DIST1' parameter of the ATNF pulsar catalogue.} $d$, and transverse proper motions $\mu_\perp$ were taken from the catalogue. For the $\gamma$-ray MSPs J1741+1351 and J1902$-$5105, no $\dot P$ value is available in the ATNF catalogue, and we have thus used the values reported in \citet{Fermi2PC}. For pulsars with known transverse proper motions, we corrected the $\dot P$ values from the kinematic Shklovskii effect \citep{shk70}, which acts to make the observed $\dot P$ values greater than the intrinsic ones by $\sim\!2.43 \times 10^{-21} \mathrm{s}^{-1}\;P_{\rm ms}\,d_{\rm pc}\,\mu_\perp^2$, where $P_{\rm ms}$ is the MSP spin period in ms, $d_{\rm pc}$ is its distance in pc, and $\mu_\perp$ is its proper motion in arcsec\,yr$^{-1}$. In a few cases, the corrections exceeded the observed $\dot P$ values, leading to negative corrected spin-down rates. For these pulsars, we kept the uncorrected $\dot P$ values, keeping in mind that the intrinsic spin-down rates are likely to be significantly smaller. The values of $P$ and $\dot P$ were then used to compute spin-down luminosities $\dot E = 4 \upi^2 I \dot P / P^3$, and the quantity $\dot E / d^2$, which can be seen as a figure of merit for $\gamma$-ray detectability. We also compiled values of the pulsar mass, $M_\mathrm{NS}$, and of $\sin i$, available in the literature.

For the selected MSPs, we computed the viewing angles via the TS99 relationship, following the prescriptions described in Section~\ref{subsec: TS99} and using the measured $M_\mathrm{NS}$ values when known, or assuming pulsar masses of $1.53 \pm 0.21$ M$_{\sun}$ \citep[this value corresponds to the average mass observed for Galactic disc MSPs with He~WD companions; for a recent compilation see Table~4 of][]{tlk12}. We note that in a few cases, namely for PSRs J1125$-$6014, J1400$-$1438, J1709+2313, J1811$-$2405, J1933$-$6211, and J2215+5135, no solutions were found for the inclination angle $i$ assuming a pulsar mass of 1.53 M$_{\sun}$, indicating that these pulsars could either have smaller masses or could originate from intermediate-mass X-ray binaries for which the $M_{\rm WD}$--$P_{\rm orb}$ relation does not apply. We discarded these pulsars from our analysis. The selected pulsars and the associated measured and derived quantities are listed in Table~\ref{table: data}.

As mentioned in the introduction, the viewing angle $\zeta$ can be estimated by fitting the radio and $\gamma$-ray profiles of MSPs in the context of theoretical models of emission from pulsars. Numerous modelling efforts have been conducted, using a variety of emission geometries \citep[e.g.,][]{Venter2009,Du2010,Johnson2012,Petri2011,Venter2012}. Detailed comparisons of theoretical treatments and results between these studies are beyond the scope of this paper. However, among published modelling studies, \citet{Johnson2012} has the largest number of consistently analysed radio and $\gamma$-ray pulsars, and therefore, for simplicity we use the preferred $\zeta_\mathrm{LC\ Modelling}$ values obtained from this study. We complemented the list of $\zeta_\mathrm{LC\ Modelling}$ values quoted in Table~\ref{table: data} by adding the results of other recent modelling analyses using similar models \citep[e.g.][]{gfc+12}, for pulsars not covered by \citet{Johnson2012}.

Finally, we completed the list of MSPs with viewing angle estimates in Table~\ref{table: data} by adding other MSPs with constraints on $\sin i$ or $\zeta_\mathrm{LC\ Modelling}$ (i.e. MSPs without an He~WD companion). Examples include PSR J0737$-$3039A in the double pulsar system, for which \citet{ksm+06} put strong constraints on the orbital inclination angle, $i$, through radio pulsar timing observations, and the isolated MSP J0030+0451, for which radio and $\gamma$-ray light-curve modelling analyses constrained the viewing angle, $\zeta$ \citep[e.g.][]{Johnson2012}.

\begin{figure}
\centering
\includegraphics[width=0.47\textwidth]{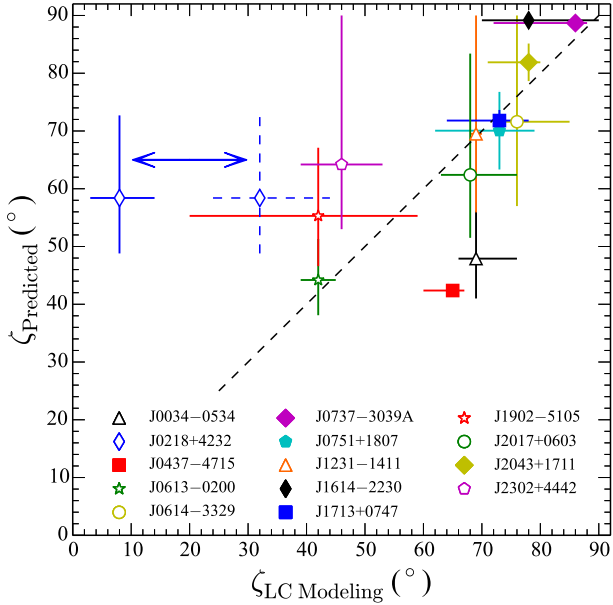}
\caption[]{Predicted viewing angles as a function of the values determined from the joint modelling of radio and $\gamma$-ray light curves, for MSPs detected in both radio and $\gamma$-rays. The filled symbols indicate MSPs with directly measured constraints on the orbital inclination angle, $i$, and open symbols represent predictions based on the TS99~relation. See Section~\ref{subsec: zeta_calc} for additional details on the determination of $\zeta_\mathrm{Predicted}$ and $\zeta_\mathrm{LC\ Modelling}$. A discussion of the two solutions shown for PSR~J0218+4232 is given in Section~\ref{subsec: 0218}.}
\label{fig: TS99_sini}
\end{figure}

A plot of the predicted viewing angles, $\zeta_\mathrm{Predicted}$, as a function of the viewing angles determined from the modelling of radio and $\gamma$-ray pulse profiles, $\zeta_\mathrm{LC\ Modelling}$, is shown in Fig.~\ref{fig: TS99_sini}. For pulsars with $\sin i$ constraints, we used the measured orbital inclination angle as the viewing angle prediction. For other pulsars, the angle estimated by the TS99~relation was used. With the exception of PSR~J0218+4232, for which $\zeta_\mathrm{LC\ Modelling}$ and $\zeta_\mathrm{Predicted}$ are markedly different (this specific MSP will be discussed later, cf. Section~\ref{subsec: 0218}), the predicted and modelled viewing angles appear to be in good agreement. Excluding PSR~J0218+4232, we find an average difference between the two angles of $\langle | \zeta_\mathrm{LC\ Modelling} - \zeta_\mathrm{Predicted} | \rangle \simeq 8^\circ$, with an rms of $\sim 8^\circ$. Furthermore, we find a Spearman rank coefficient for this data set of $r_s \sim 0.88$, close to 1, and therefore indeed suggesting a positive correlation between $\zeta_\mathrm{LC\ Modelling}$ and $\zeta_\mathrm{Predicted}$ values. A probability of chance correlation of only $8 \times 10^{-5}$ is found for this value of $r_s$. Including PSR~J0218+4232, the average difference becomes $11^\circ$ with an rms of $13^\circ$ and therefore also consistent with 0. We conclude that the procedure described in this section provides a reliable method for estimating the true viewing angle of a pulsar. In the following, we use the $\zeta$ angles determined for the selected sample of MSPs to search for a possible relation between MSP viewing angles and $\gamma$-ray detectability.


\section{Results}

\subsection{The distribution of MSP viewing angles} 
\label{subsec: viewing}

We assign viewing angles, $\zeta$, to the MSPs in our sample in the following priority order:

\begin{itemize}
\item for MSPs with viewing angle constraints obtained from light-curve modelling analyses, the best-fitting $\zeta$ value is used, \\

\item if $\zeta$ constraints from light-curve modelling studies do not exist, but the quantity $\sin i$ has been measured from e.g. determination of the Shapiro delay or optical observations of the companion, we use $i$ for the viewing angle, assuming that it is equivalent to $\zeta$ in these systems, as discussed in Section~\ref{subsec: TS99}, \\

\item if the above measurements do not exist, we calculate $\zeta$ using the TS99 relation, as described in Section~\ref{subsec: TS99}.
\end{itemize}

\begin{figure}
\centering
\includegraphics[width=0.47\textwidth]{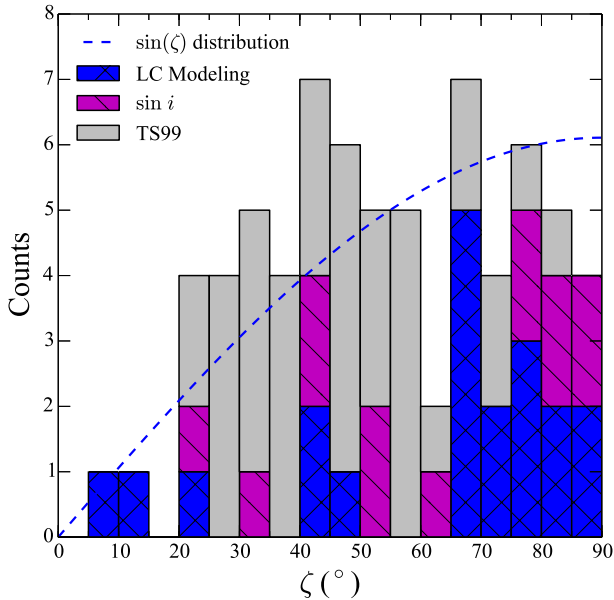}
\caption[]{Distribution of viewing angles, $\zeta$, for MSPs with radio and $\gamma$-ray light-curve modelling constraints (blue, cross-hatched), with orbital inclination measurements (purple, hatched), or with viewing angle predictions based only on the TS99~relation between $M_{\rm WD}$ and $P_{\rm orb}$ (grey shaded). See Section~\ref{subsec: viewing} for details on the construction of the histogram. The dashed blue line indicates the expectation of an isotropic distribution of viewing angles.} 
\label{fig: viewing}
\end{figure}

Fig.~\ref{fig: viewing} displays the distribution of viewing angles for our total sample of 70 MSPs. A $\chi^2$ test shows that the derived distribution is consistent with that expected for an isotropic (sine-like) distribution at the 85 per cent confidence level, which supports our trust in the applied methodology presented here in determining the $\zeta$ values.

It is interesting to note that the distribution of viewing angles shown in Fig.~\ref{fig: viewing} does not appear to favour any particular direction. An anisotropy in the viewing angle distribution could for example result from the alignment of the magnetic axis with the spin axis \citep[][and references therein]{tm98}. Thus, assuming that radio emission beams are produced above the magnetic poles and have a certain width would make pulsars seen under large viewing angles unlikely to be detected if the angles, $\alpha$, between the spin axis and the magnetic axis were preferentially closer to 0. With an isotropic distribution of viewing angles, we could infer that the distribution of $\alpha$ angles is also isotropic for this sample, or only weakly directional. This conclusion must at present be taken with a grain of salt as the number of pulsars in the sample is very limited.

\subsection{$\gamma$-ray detectability of radio MSPs} 
\label{subsec: detectability}

\begin{figure}
\centering
\includegraphics[width=0.47\textwidth]{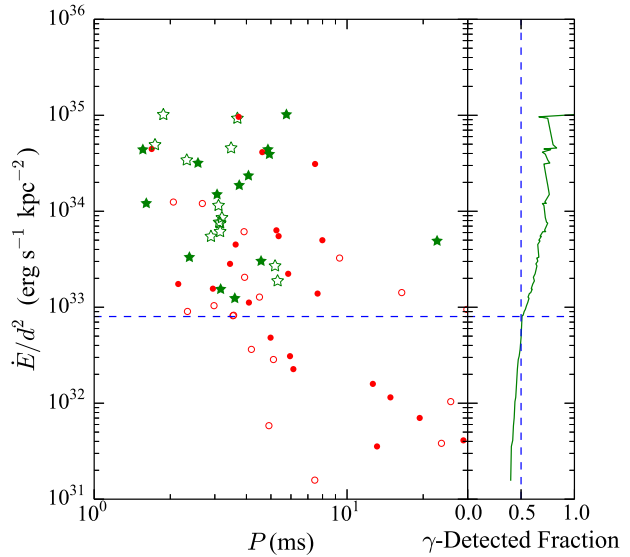}
\caption[]{Spin-down luminosity, $\dot E$, normalized by the square of the distance, $d$, as a function of the rotational period, $P$, for the sample of MSPs selected in Section \ref{subsec: zeta_calc}. MSPs detected as pulsed sources of $\gamma$-rays are shown as green stars, while red circles represent undetected ones. Whenever possible (filled symbols), the $\dot E$ values were corrected for the Shklovskii effect. The right-hand panel shows the fraction of $\gamma$-ray-detected MSPs in the sample, as a function of decreasing values of $\dot E / d^2$. The horizontal dashed blue line, at $\dot E / d^2 = 8 \times 10^{32}$ erg~s$^{-1}$ kpc$^{-2}$, indicates the limit above which 50 per cent of MSPs in this sample are detected in $\gamma$-rays.}
\label{fig: Edotd2_P0}
\end{figure}

In Fig.~\ref{fig: Edotd2_P0}, we show a plot of $\dot E / d^2$ as a function of the rotational period for the MSPs in our sample. As expected, $\gamma$-ray detected MSPs occupy the upper part of the plot which confirms that $\dot E / d^2$ is a good measure of potential detectability in $\gamma$-rays. Two thirds of the MSPs in our sample with $\dot E / d^2$ values larger than $\simeq 3 \times 10^{33}\;{\rm erg}\,{\rm s}^{-1}\,{\rm kpc}^{-2}$ are detected in $\gamma$-rays, and the 50 per cent detection level corresponds to $\dot{E}/d^2 \simeq 8 \times 10^{32}\;{\rm erg}\,{\rm s}^{-1}\,{\rm kpc}^{-2}$. It should be noted that nearly all MSPs in our sample with $\dot{E}/d^2$ values above $8 \times 10^{32}\;{\rm erg}\,{\rm s}^{-1}\,{\rm kpc}^{-2}$ have spin-down luminosities $\dot E$ larger than a few times $10^{33}\;{\rm erg}\,{\rm s}^{-1}$ and are therefore potential $\gamma$-ray emitters: the least energetic $\gamma$-ray MSPs detected to date by the LAT have $\dot E$ values close to $10^{33}\;{\rm erg}\,{\rm s}^{-1}$ \citep[see e.g. fig. 9 of][]{Fermi2PC}. These sources with $\dot{E}/d^2 \geq 8 \times 10^{32}\;{\rm erg}\,{\rm s}^{-1}\,{\rm kpc}^{-2}$ can therefore be used to identify the causes of non-detection of certain MSPs in $\gamma$-rays.

\begin{figure*}
\centering
\includegraphics[width=0.8\textwidth]{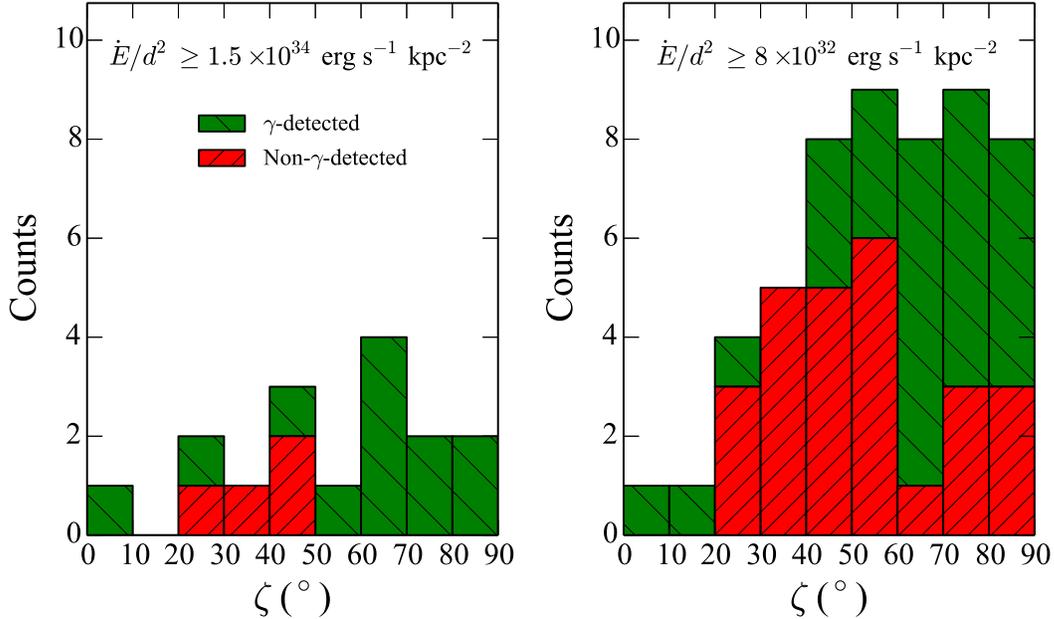}
\caption[]{The distribution of viewing angles, $\zeta$, for $\dot{E}/d^2 \ge 1.5\times 10^{34}\;{\rm erg}\,{\rm s}^{-1}\,{\rm kpc}^{-2}$ (left-hand panel) and for $\dot{E}/d^2 \ge 8 \times 10^{32}\;{\rm erg}\,{\rm s}^{-1}\,{\rm kpc}^{-2}$ (right-hand panel). The $\gamma$-detected and undetected MSPs are marked in green and red, respectively. It is noted that the ratio of $\gamma$-ray detected-to-undetected MSPs is about 3:1 for the very energetic, nearby MSPs in the left-hand panel and about 1:1 for MSPs with smaller values of $\dot{E}/d^2$ in the right-hand panel.}
\label{fig: histograms}
\end{figure*}

In Fig.~\ref{fig: histograms}, we show the distributions of viewing angles for MSPs with $\dot E / d^2 \ge 1.5 \times 10^{34}\;{\rm erg}\,{\rm s}^{-1}\,{\rm kpc}^{-2}$ and with $\dot E / d^2 \ge 8 \times 10^{32}\;{\rm erg}\,{\rm s}^{-1}\,{\rm kpc}^{-2}$, respectively. In the former sample, 75 per cent of the pulsars are seen in $\gamma$-rays, while half of the objects in the second sample are $\gamma$-ray pulsars. We denote the two MSP samples as $S_{75}$ and $S_{50}$, respectively. Taken together, the distributions of $\zeta$ angles for $\gamma$-undetected and $\gamma$-detected pulsars do not deviate much from sine-like, isotropic distributions. Nevertheless, MSPs that are not detected in $\gamma$-rays appear to be distributed, on average, towards smaller $\zeta$ values compared to the detected ones: 38$^\circ$ versus 59$^\circ$ for sample $S_{75}$ and 53$^\circ$ versus 62$^\circ$ for sample $S_{50}$. A one-dimensional Kolmogorov--Smirnov (KS) test \citep{Press1992} indicates that the probability that the $\gamma$-detected and non-detected ones originate from the same parent distribution is only about 3 per cent for each of the two MSP samples. We note that the conclusion is the same when only considering our predicted $\zeta$ values and thus not using the $\zeta$ values obtained from light-curve modelling studies: the average $\zeta$ values for the various distributions are consistent with the ones listed above to within $3^\circ$ in all cases, and the KS test again indicates a small probability of about 5 per cent that the $\gamma$-detected and non-detected pulsars originate from the same parent distribution for both samples.

As can be seen from the modelling presented in e.g. \citet{Venter2009} or \citet{Takata2011} for the outer gap and the two-pole caustic emission geometries, $\gamma$-ray detectability is determined by the combination of the magnetic inclination angle, $\alpha$, and the viewing angle, $\zeta$. Pulsars with small viewing angles or small magnetic inclination angles are often not expected to produce detectable $\gamma$-ray emission, either because of their emission beams not crossing the line of sight to the Earth or as a result of weak modulation in the emission making them difficult to detect. In contrast, under configurations with large $\alpha$ or $\zeta$ values, one generally expects marked $\gamma$-ray signal modulation with sharp peaks that are easily detected. The present observational analysis supports these theoretical hypotheses: we find that there is a slight difference in the viewing angle distributions for the $\gamma$-ray MSPs and the non-detected MSPs, although only at a marginally significant level, such that MSPs that are non-detected in $\gamma$-rays tend to have smaller viewing angles in general. Therefore, we conclude that small values of the viewing angle $\zeta$ are at least partly responsible for the non-detection of MSPs with large values of $\dot E / d^2$. Furthermore, we postulate that small values of $\alpha$ and/or the possibility of overestimated values of $\dot{E} / d^2$ play a role as well.

Our method for estimating the angles under which the MSPs in binary systems are likely seen may be even more useful in the future when significantly more pulsars are detected in the radio and the $\gamma$-ray domains. Thereby, this method, together with simple statistical arguments based on beaming directions with respect to the line of sight, can hopefully help to understand the causes of the non-detection of some energetic pulsars in $\gamma$-rays.

\subsection{PSR~J0218+4232}
\label{subsec: 0218}

As noted from Fig.~\ref{fig: TS99_sini}, PSR~J0218+4232 \citep{ndf+95} is quite an outlier with respect to the $\zeta _{\rm Predicted}=\zeta _{\rm LC\,Modelling}$ diagonal. The predicted value from the TS99~relation is roughly $\zeta _{\rm TS99}\approx 60^{\circ} \pm 10^{\circ}$, depending on the assumed pulsar mass, $M_{\rm NS}$ (see Fig.~\ref{fig:J0218}). The most likely viewing angle suggested by \citet{Johnson2012} from modelling of radio and $\gamma$-ray pulse profiles is much smaller, about $\zeta _{\rm LC\,Modelling}\approx 8^{\circ} \pm 6^{\circ}$. However, from the measured mass function of the radio pulsar, it is highly unlikely that the orbital inclination angle is that small. From a statistical point of view, the probability is small: a random (isotropic) distribution of orbital inclination angles would yield a probability of less than 1 per cent for the case that $i\le 8^{\circ}$ (and $\le 3$ per cent for $i\le 14^{\circ}$). More importantly, such a small inclination angle of $i=8^{\circ}$ would imply a WD mass of $1.9\;M_{\sun}$ (or $M_{\rm WD}=0.81\;M_{\sun}$ for $i=14^{\circ}$) even for the smallest value of $M_{\rm NS}$ (here assumed to be $1.11\;M_{\sun}$). Such a high WD mass is unrealistic and is clearly in contradiction with the optical observations by \citet{bvk03}, who find evidence for a $\sim\!0.2\;M_{\sun}$ He~WD companion, in nice agreement with TS99. A low-mass He WD ($\sim\!0.2\;M_{\sun}$) requires $i>45^{\circ}$ for $M_{\rm NS} > 1.11\;M_{\sun}$. It is worth noting that \citet{Johnson2012} finds an alternative solution of $\zeta _{\rm LC\,Modelling}$ under the outer gap model at $32_{-8}^{+12}\,^{\circ}$, whose $1\,\sigma$ upper limit is just in agreement with this requirement on $i$ (see the dashed alternative solution for PSR~J0218+4232 in Fig.~\ref{fig: TS99_sini}).

\subsection{PSR~J0034$-$0534}

PSR~J0034$-$0534 \citep{bhl+94} is an MSP orbiting a WD with an orbital period of $P_{\rm orb}=1.59\;{\rm d}$. The TS99~relation predicts $0.202 < M_{\rm WD}/M_{\sun} < 0.220$ which results in $i=\zeta _{\rm TS99}\approx 48^{\circ} \pm 8^{\circ}$ (the actual $1\,\sigma$ error bars are slightly asymmetric, see Table~\ref{table: data}) if $M_{\rm NS}= 1.53\pm 0.21\;M_{\sun}$. The radio and $\gamma$-ray light-curve modelling yields $\zeta _{\rm LC\,Modelling}\approx 69^{+7}_{-3}\,^{\circ}$. If indeed $i\ge 69^{\circ}$, it would require a massive pulsar of $M_{\rm NS}>2.0\;M_{\sun}$, according to the TS99~relation. PSR~J0034$-$0534 is therefore an interesting target for a future precise $M_{\rm NS}$ measurement.

\subsection{PSR~B1855+09 (J1857+0943)}

This pulsar has a measured mass of $1.57^{+0.12}_{-0.11}\;M_{\sun}$, according to \citet{nss03}. Using the TS99~relation, we constrain this 12.3~d binary MSP to have an He~WD companion with a mass of $M_{\rm WD}<0.275\;M_{\sun}$, which yields an upper limit on the pulsar mass of $1.54\;M_{\sun}$ (for $i=90^{\circ}$). Future timing and improvements of its Shapiro delay measurement may confirm this $M_{\rm NS}$ limit.

\subsection{PSR~J1327$-$0755}

While checking for kinematic corrections of the MSPs presented in Table~1, we noticed a very significant discrepancy for the estimated distance to PSR~J1327$-$0755. This source was recently discovered by \citet{blr+13} and has an estimated dispersion measure distance of $d\simeq 1.7\;{\rm kpc}$ which, according to the authors, could be an overestimate when applying the NE2001 model for the Galactic electron density distribution along a line of sight off the Galactic plane. The pulsar has a large measured proper motion of $99\pm 47 \;{\rm mas\,yr}^{-1}$, which we find must require a huge reduction in its true distance by a factor of $\sim\!6$. Otherwise, when correcting for kinematic effects due to the Shklovskii~effect, as well as vertical and differential rotational acceleration in our Galaxy \citep[e.g. using an expression analogous to equation 16 in][]{lwj+09}, we find a negative value for the intrinsic $\dot{P}$, which is not possible. Alternatively, the discrepancy is solved if the proper motion, $\mu$, is smaller by a factor of 2.4 (which is not impossible given the large error bar on this number).


\section{On the B-fields of $\gamma$-ray MSPs}

Among the population of $\gamma$-ray MSPs, there is a handful of sources that are very efficient at converting spin-down luminosity into emission, i.e. they have large $\eta = L_\gamma / \dot E$ values, where $\eta$ denotes the $\gamma$-ray efficiency and $L_\gamma$ is the $\gamma$-ray luminosity. For rotation-powered pulsars (where no other energy sources are available), only values of $\eta \leq 1$ are possible. Note, magnetars have X-ray luminosities exceeding the loss rate of rotational energy ($\eta _x = L_x/\dot{E}_{\rm rot}>1$). This is possible if the emission is powered by the instability and decay of their strong B-fields \citep{td95}. For the $\gamma$-ray emitting MSPs, however, this is not possible given their weak B-fields. Hence, their $\gamma$-ray emission must be powered by spin-down energy. For some MSPs, the $\eta$ parameter is found to be close to 100 per cent or larger \citep{Fermi2PC}. However, overestimated distances could be responsible for some of the large $\eta$ values. For example, as discussed in \citet{Fermi2PC} and \citet{Espinoza2013}, material in the direction of PSR~J0610$-$2100 unaccounted for in the NE2001 model could make its predicted distance significantly smaller, which would in turn decrease its currently (unrealistic) very large $\gamma$-ray efficiency of $\sim 1200$ per cent. Nevertheless, it is still possible that some of the MSPs are genuinely very efficient at converting their energy budget into $\gamma$-ray emission. In this section, we investigate the physical properties of MSPs with large $\eta$ values approaching 1 and demonstrate that their surface B-fields can be much weaker than derived from the classical dipole formula.

The loss of rotational energy of a non-accreting MSP is caused by a combination of magnetic dipole radiation \citep{pac67}, the presence of plasma currents in the magnetosphere \citep[the Goldreich-Julian term,][]{gj69} and gravitational wave radiation \citep[e.g.][]{wsk+13}:

\begin{equation}
\dot{E}_{\rm rot} = \dot{E}_{\rm dipole} + \dot{E}_{\rm GJ}  + \dot{E}_{\rm gw}, 
\label{eq:Edot}
\end{equation}
where 
\begin{equation}
\dot{E}_{\rm dipole} = -\frac{2}{3 c^3}|\ddot{\bmath{m}}|^2\qquad \wedge \qquad |\ddot{\bmath{m}}|=B_0R^3\Omega^2\sin\alpha,
\label{eq:Edipole}
\end{equation}
and $\dot{E}_{\rm GJ}$ is found by considering the outward Poynting energy flux $S\sim c B^2/4\upi$ crossing the light cylinder $r_{\rm lc}$, giving rise to the observed high-frequency radiation as well as emission of relativistic particles. Assuming a dipolar form of the B-field within this zone, i.e. $B_{\rm lc}\propto B_0(R/r_{\rm lc})^3$, one finds the well-known expression 
\begin{equation}
|\dot{E}_{\rm GJ}| \sim 4\upi r_{\rm lc}^2\,S \sim B_0^2R^6\Omega^4/c^3,
\label{eq:Eplasma}
\end{equation}
where $B_0$ is the magnetic flux density at the surface of the neutron star, $R$ is its radius, $\Omega=2\upi/P$ is its spin angular velocity, $\alpha$ is the magnetic inclination angle and $\bmath{m}$ is the magnetic moment of the neutron star. In the following, we assume $\dot{E}_{\rm gw}\ll \dot{E}_{\rm rot}$ \citep{aaa+10} and disregard the third term in equation~(\ref{eq:Edot}). We also disregard a recently suggested term related to quantum vacuum friction \citep{drb12}.

The dipole component of pulsar B-fields is traditionally found simply by equating the loss rate of rotational energy ($\dot{E}_{\rm rot}=-4\upi^2I\dot{P}/P^3$) to the energy loss caused by magnetic dipole radiation ($\dot{E}_{\rm dipole}$):
\begin{equation}
B_{\rm dipole}=C \cdot \sqrt{P\dot{P}}, 
\label{eq:B-dipole}
\end{equation}
where the constant is taken to be $C=3.2\times10^{19}\;{\rm G\,s}^{-1/2}$ for the equatorial B-field strength (assuming $R=10\;{\rm km}$, $I=10^{45}\;{\rm g\,cm}^2$, and $\alpha=90^{\circ}$)\footnote{After the discovery of intermittent pulsars \citep{klo+06}, there have been attempts in the literature to take into account the $\bmath{j}\times \bmath{B}$ force exerted by plasma currents in the magnetosphere to yield a combined spin-down torque \citep[e.g.][]{spi06} and thus a revision of the expression for $B_0$ \citep[e.g.][]{tlk12}.}.

Defining $x$ and $y=1-x$ in the following manner
\begin{equation}
\dot{E}_{\rm dipole} \equiv x\cdot \dot{E}_{\rm rot} \qquad \wedge \qquad
\dot{E}_{\rm GJ} \equiv y\cdot \dot{E}_{\rm rot} 
\label{eq:xy}
\end{equation}
and introducing the parameters $\kappa$ and $\eta$
\begin{equation}
L_{\gamma} \equiv \kappa \cdot |\dot{E}_{\rm GJ}| = \kappa y \cdot |\dot{E}_{\rm rot}| \equiv \eta \cdot |\dot{E}_{\rm rot}|
\label{eq:xyetagamma}
\end{equation}
allows us to write
\begin{equation}
\dot{E}_{\rm dipole}=\left( 1-\frac{\eta}{\kappa} \right)\,\dot{E}_{\rm rot}.
\label{eq:dipole_eta_kappa}
\end{equation}
This leads to a revised expression for the surface B-field strength of the pulsar:
\begin{equation}
B_\star=C \cdot \sqrt{\left(1-\frac{\eta}{\kappa}\right) P\dot{P}}. 
\label{eq:B-dipole_revised}
\end{equation}

\begin{figure}
\centering
\includegraphics[angle=-90,width=0.47\textwidth]{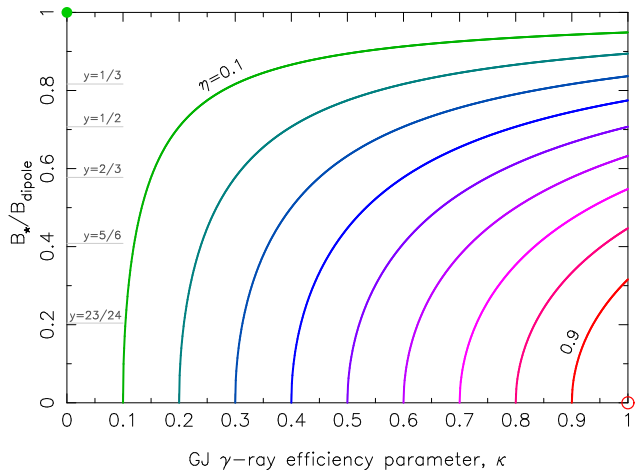}
\caption{The reduction factor of the derived B-field strengths in units of the classical, pure dipole field spin-down estimate, plotted as a function of the efficiency parameters $\kappa$ and $\eta$, defined in equation~(\ref{eq:xyetagamma}). The various curves are plotted for $0.1 \le \eta \le 0.9$, in steps of 0.1. The $B_\star /B_{\rm dipole}$ ratio is marked for various values of $y=\eta / \kappa$. The filled green circle at the upper-left corner is for a pure dipole torque. The open red circle at the lower-right corner is unphysical ($\eta=\kappa=1$).}
\label{fig:Fermi2}
\end{figure}

The conclusion from equation~(\ref{eq:B-dipole_revised}) is that pulsars which are very efficient at converting rotational energy into $\gamma$-rays (large values of $\eta$) may have $B_\star \ll B_{\rm dipole}$. Even pulsars with small values of $\eta$ could also have B-fields which are significantly smaller than the value derived from the classical formula in equation~(\ref{eq:B-dipole}) if their $\gamma$-ray emission is inefficient ($\eta \le \kappa \ll 1$) despite a large value of $\dot{E}_{\rm GJ}$. In Fig.~\ref{fig:Fermi2}, we illustrate this point by plotting $B_\star/B_{\rm dipole}$ as a function of $\eta$ and $\kappa$. MSPs with $\eta$ values close to 1 may thus have considerably smaller B-fields than inferred from the classical formula, e.g. by a factor of 2--5 (in principle, possibly even more), see Fig.~\ref{fig:Fermi2}.

Although the dipole torque dependence on the magnetic inclination angle, $\alpha$, is cancelled out when considering the ratio $B_\star /B_{\rm dipole}$, we notice that the relative proportion of $\dot{E}_{\rm dipole}$ and $\dot{E}_{\rm GJ}$ to the total $\dot{E}_{\rm rot}$ (i.e. $x$ and $y$) depends on $\alpha$ \citep[e.g.][]{spi06,kkhc12,lst12}. Hence, naively, one might expect less efficient production of $\gamma$-rays for more orthogonal rotators, where the relative importance of the plasma term is smaller, thus resulting in smaller values of $L_\gamma$ for pulsars with a large viewing angle, $\zeta$. The reason is that, on average for a large sample of pulsars, $\zeta$ is expected to trace $\alpha$ for a given distribution of impact parameters, $\beta$, within the beam boundary. However, the picture is more complicated, not only in terms of the different morphology for the emitting radio and $\gamma$-ray beams but also in relation to the interesting related question of the efficiency of the torque produced by the currents as $\eta \rightarrow 1$. The answer seems to depend on where the dissipation into $\gamma$-rays is occurring. If it happens in the current sheet outside the light cylinder, the torque may be larger than if the emission occurs well inside the light cylinder \citep[e.g.][and references therein]{kkhc12}.

The increasing number of intermittent pulsars \citep{ksm+06,Lorimer2012,Young2013} may enable the possibility to set constraints on $x$ and $y\equiv \eta$/$\kappa$ based on measurements of the different spin-down torques acting on pulsars in their {\em on} and {\em off} states, respectively. Although the phenomenon has not yet been observed for MSPs, there may be possibilities to estimate the reduced B-fields from equation~(\ref{eq:B-dipole_revised}) in the near future. We strongly encourage further investigations of MSP emission properties, the energy budget involved for the different torques in action, as well as the structure and surface strength of the MSP B-fields. 


\section{Summary}
\label{sec:summary}

\begin{itemize}

\item We have demonstrated that the viewing angles of binary MSPs (as inferred from $\gamma$-ray light-curve modelling) are well described by their orbital inclination angles (estimated from the $M_{\rm WD}$--$P_{\rm orb}$ relation), which confirms that pulsar spin axes do indeed align with the orbital momentum vector during recycling. This has been an important assumption in LMXB modelling for many years, but hitherto not verified from a systematic investigation of viewing angles and orbital inclinations.

\item From our extended sample of predicted viewing angles, in complement with values obtained from other constraints when available (yielding a total of 70~MSPs), we have been able to study the non-detection of energetic and nearby MSPs in view of beaming geometry and orientation. We find evidence for slightly different viewing angle distributions for $\gamma$-ray detected and non-detected MSPs. Although marginally significant, this result suggests that energetic and nearby MSPs are mainly undetected in $\gamma$-rays simply because they are seen under unfavourable (small) viewing angles, such that their emission beams do not cross the line of sight to the Earth, or the modulation of the $\gamma$-ray emission is very limited.

\item We have discussed the B-fields of pulsars with high $\gamma$-ray luminosities and pointed out that pulsars that are efficient at converting their rotational energy into $\gamma$-ray emission may have significantly overestimated dipolar B-field strengths. We encourage further studies on this issue.

\end{itemize}


\section*{Acknowledgements}

We thank Axel Jessner, Alice K. Harding, David A. Smith, Tyrel J. Johnson, Richard N. Manchester, and Ramesh Karuppusamy for discussions or comments on the manuscript. We thank the anonymous referee for constructive suggestions. TMT gratefully acknowledges financial support and hospitality at both the Argelander-Institut f\"ur Astronomie, Universit\"at Bonn and the Max-Planck-Institut f\"ur Radioastronomie. We made extensive use of the ATNF pulsar catalogue \citep{mhth05}. We also made extensive use of the list of neutron star masses assembled by James M. Lattimer and Andrew W. Steiner\footnote{http://stellarcollapse.org/nsmasses} \citep{Lattimer2012}.

\bibliographystyle{mnras} 
\bibliography{GT13}

\clearpage

\onecolumn
\begin{landscape}
\begin{center}
\begin{tiny}
\begin{longtable}{p{1.5cm}cccccccccccccccc}

\caption{Properties of radio MSPs likely to be orbiting He~WD companions, according to the criteria defined by the appendix in \citet{tlk12}. Also included are MSPs with direct observational constraints on their orbital inclination, or with viewing angle constraints obtained from the joint modelling of radio and $\gamma$-ray emission profiles. For each pulsar, we quote the rotational period $P$, the observed spin-down rate $\dot P$, and the orbital period $P_\mathrm{orb}$ and projected semimajor axis $a_\mathrm{p} \sin i$ for pulsars in binary systems. Distances $d$ are taken from the `DIST1' column of the ATNF pulsar catalogue, as well as transverse proper motions $\mu_\perp$. For pulsars with known $\mu_\perp$ values, the spin-down luminosities $\dot E$ and $\dot E / d^2$ quantities were corrected for the Shklovskii effect. In some cases, the corrections exceeded the values themselves; for these pulsars, we give the uncorrected $\dot E$ and $\dot E / d^2$ values, and mark them with a $\dag$ symbol. The `$\gamma$' column indicates whether the pulsar is detected in GeV $\gamma$-rays, and the `$\eta$' column reports the measured efficiencies of conversion of $\dot E$ into $\gamma$-ray emission (the assumed $\dot E$ and distance values may differ from the ones given in this table). Measured neutron star masses and orbital inclination constraints are given in the $M_\mathrm{NS}$ and $\sin i$ columns. The following two columns list viewing angle constraints obtained from the modelling of radio and $\gamma$-ray light curves in the context of geometrical models of emission from MSPs and the corresponding model assumed. The `$\zeta_\mathrm{TS99}$' column gives the viewing angles as expected from the TS99~relation, using the measured neutron star mass when available (values marked with a $\bigstar$ symbol) or assuming a mass of 1.53 M$_{\sun}$. The last column lists the references for the quoted $\eta$, $M_\mathrm{NS}$, $\sin i$, and $\zeta_\mathrm{LC\ Modelling}$ values: (1) -- \citet{Fermi2PC}, (2) -- \citet{Guillemot2013}, (3) -- \citet{Hou2013}, (4) -- \citet{Kaplan2012}, (5) -- \citet{vbv+08}, (6) -- \citet{nsk08}, (7) -- \citet{ksm+06}, (8) -- \citet{lcw+01}, (9) -- \citet{Deller2012}, (10) -- \citet{Thorsett1999}, (11) -- \citet{dpr+10}, (12) -- \citet{sns+05}, (13) -- \citet{akk+12}, (14) -- \citet{fsk+10}, (15) -- \citet{gsf+11}, (16) -- \citet{nss03}, (17) -- \citet{fbw+11}, (18) -- \citet{hbo06}, (19) -- \citet{dfc+12}, (20) -- \citet{nss01}, (21) -- \citet{gfc+12}, (22) -- \citet{Kasian2012}, (23) -- \citet{Callanan1998}, (24) -- \citet{vbc+09}, (25) -- \citet{llww05}, and (26) -- \citet{Johnson2012}.
\label{table: data}}\\

\hline 
Pulsar & $P$ & $\dot P$ & $P_\mathrm{orb}$ & $a_\mathrm{p} \sin i$ & $d$ & $\mu_\perp$ & $\dot E$ & $\dot E / d^2$ & $\gamma$ & $\eta$ & $M_\mathrm{NS}$ & $\sin i$ & $\zeta_\mathrm{LC\ Modelling}$ & Model & $\zeta_\mathrm{TS99}$ & Refs \\

 & (ms) & ($10^{-20}$) & (d) & (lt-s) & (kpc) & (mas yr$^{-1}$) & ($10^{33}$ erg s$^{-1}$) & ($10^{33}$ erg s$^{-1}$ kpc$^{-2}$) & & (per cent) & (M$_{\sun}$) & & ($^\circ$) & & ($^\circ$) & \\

\hline
\endfirsthead

\multicolumn{14}{c}{{Continued from previous page.}}\\
\hline
Pulsar & $P$ & $\dot P$ & $P_\mathrm{orb}$ & $a_\mathrm{p} \sin i$ & $d$ & $\mu_\perp$ & $\dot E$ & $\dot E / d^2$ & $\gamma$ & $\eta$ & $M_\mathrm{NS}$ & $\sin i$ & $\zeta_\mathrm{LC\ Modelling}$ & Model & $\zeta_\mathrm{TS99}$ & Refs \\

 & (ms) & ($10^{-20}$) & (d) & (lt-s) & (kpc) & (mas yr$^{-1}$) & ($10^{33}$ erg s$^{-1}$) & ($10^{33}$ erg s$^{-1}$ kpc$^{-2}$) & & (per cent) & (M$_{\sun}$) & & ($^\circ$) & & ($^\circ$) & \\

\hline
\endhead

\hline
\endfoot

\hline
\endlastfoot

J0030+0451 & 4.87 & 1.02 & -- & -- & 0.3 & 5.7 & 3.5 & 44.0 & Y & 16 & -- & -- & $66_{-2}^{+4}$ & OG & -- & 1, --, --, 26\\
J0034$-$0534 & 1.88 & 0.50 & 1.6 & 1.4 & 0.5 & -- & 29.6 & 101.6 & Y & 3.3 & -- & -- & $69_{-3}^{+7}$ & alTPC & $47.9_{-6.9}^{+8.3}$ & 1, --, --, 26\\
J0101$-$6422 & 2.57 & 0.52 & 1.8 & 1.7 & 0.6 & 15.6 & 10.0 & 31.8 & Y & 3.8 & -- & -- & -- & -- & $53.5_{-8.2}^{+10.7}$ & 1, --, --, --\\
J0218+4232 & 2.32 & 7.74 & 2.0 & 2.0 & 2.7 & -- & 243.7 & 34.2 & Y & 16 & -- & -- & $8_{-5}^{+6}$ & TPC & $58.4_{-9.6}^{+14.3}$ & 1, --, --, 26\\
J0407+1607 & 25.70 & 7.90 & 669.1 & 106.5 & 1.3 & -- & 0.2 & 0.1 & -- & -- & -- & -- & -- & -- & $29.6_{-3.4}^{+3.8}$ & --, --, --, --\\
J0437$-$4715 & 5.76 & 5.73 & 5.7 & 3.4 & 0.2 & 141.3 & 2.6 & 101.8 & Y & 1.7 & $1.76_{-0.20}^{+0.20}$ & 0.674(3) & $65_{-5}^{+2}$ & TPC & $45.2_{-5.8}^{+6.9}\bigstar$ & 1, 5, 5, 26\\
J0613$-$0200 & 3.06 & 0.96 & 1.2 & 1.1 & 0.9 & 10.8 & 12.1 & 15.0 & Y & 24.1 & -- & -- & $42_{-3}^{+3}$ & TPC & $44.2_{-6.1}^{+7.1}$ & 1, --, --, 26\\
J0614$-$3329 & 3.15 & 1.75 & 53.6 & 27.6 & 1.9 & -- & 22.1 & 6.1 & Y & 215 & -- & -- & $76_{-4}^{+9}$ & TPC & $71.6_{-14.6}^{+18.4}$ & 1, --, --, 26\\
J0621+1002 & 28.85 & 4.73 & 8.3 & 12.0 & 1.4 & 3.5 & $7.6 \times 10^{-2}$ & $4.1 \times 10^{-2}$ & -- & -- & $1.70_{-0.17}^{+0.10}$ & 0.4(4) & -- & -- & -- & --, 6, 22, --\\
J0737$-$3039A & 22.70 & 175.99 & 0.1 & 1.4 & 1.1 & 4.4 & 5.9 & 4.9 & Y & 10 & $1.3381_{-0.0007}^{+0.0007}$ & 0.99974(39) & $86_{-14}^{+2}$ & TPC & -- & 2, 7, 7, 2\\
J0751+1807 & 3.48 & 0.78 & 0.3 & 0.4 & 0.4 & -- & 7.3 & 45.6 & Y & 3.5 & $1.26_{-0.14}^{+0.14}$ & 0.94(4) & $73_{-11}^{+6}$ & OG & $44.3_{-5.0}^{+5.5}\bigstar$ & 1, 6, 6, 26\\
J1012+5307 & 5.26 & 1.71 & 0.6 & 0.6 & 0.7 & 25.3 & 3.1 & 6.3 & -- & -- & $1.64_{-0.22}^{+0.22}$ & 0.78(4) & -- & -- & $40.3_{-5.3}^{+5.8}\bigstar$ & --, 8, 23, --\\
J1017$-$7156 & 2.34 & 0.26 & 6.5 & 4.8 & 3.0 & -- & 8.0 & 0.9 & -- & -- & -- & -- & -- & -- & $58.5_{-9.7}^{+15.0}$ & --, --, --, --\\
J1022+1001 & 16.45 & 4.33 & 7.8 & 16.8 & 0.5 & -- & 0.4 & 1.4 & -- & -- & -- & 0.7(1) & -- & -- & -- & --, --, 18, --\\
J1023+0038 & 1.69 & 1.20 & 0.2 & 0.3 & 1.4 & 18.0 & 83.6 & 44.5 & -- & -- & $1.71_{-0.16}^{+0.16}$ & 0.67(3) & -- & -- & $63.4_{-8.4}^{+12.3}\bigstar$ & --, 9, 9, --\\
J1045$-$4509 & 7.47 & 1.77 & 4.1 & 3.0 & 0.2 & 8.0 & 1.6 & 31.1 & -- & -- & $1.19_{-0.29}^{+0.29}$ & -- & -- & -- & $41.2_{-8.2}^{+9.8}\bigstar$ & --, 10, --, --\\
J1125$-$5825 & 3.10 & 5.96 & 76.4 & 33.6 & 2.6 & -- & 78.9 & 11.5 & Y & 9.1 & -- & -- & -- & -- & $60.9_{-10.3}^{+18.2}$ & 1, --, --, --\\
J1216$-$6410 & 3.54 & 0.16 & 4.0 & 2.9 & 1.3 & -- & 1.4 & 0.8 & -- & -- & -- & -- & -- & -- & $48.4_{-7.1}^{+8.7}$ & --, --, --, --\\
J1231$-$1411 & 3.68 & 2.28 & 1.9 & 2.0 & 0.4 & 104.4 & 18.0\dag & 93.0\dag & Y & 45.9 & -- & -- & $69_{-1}^{+1}$ & TPC & $69.5_{-13.6}^{+20.5}$ & 1, --, --, 26\\
J1327$-$0755 & 2.68 & 1.77 & 8.4 & 6.6 & 1.7 & 98.8 & 36.4\dag & 12.0\dag & -- & -- & -- & -- & -- & -- & $73.9_{-16.0}^{+16.1}$ & --, --, --, --\\
J1455$-$3330 & 7.99 & 2.43 & 76.2 & 32.4 & 0.5 & 24.5 & 1.4 & 5.0 & -- & -- & -- & -- & -- & -- & $57.4_{-9.2}^{+13.8}$ & --, --, --, --\\
J1543$-$5149 & 2.06 & 1.61 & 8.1 & 6.5 & 2.4 & -- & 73.0 & 12.5 & -- & -- & -- & -- & -- & -- & $76.1_{-17.2}^{+13.9}$ & --, --, --, --\\
J1600$-$3053 & 3.60 & 0.95 & 14.3 & 8.8 & 2.4 & 7.2 & 7.1 & 1.2 & Y & 23 & -- & 0.8(2) & -- & -- & $57.6_{-9.5}^{+14.3}$ & 1, --, 24, --\\
J1603$-$7202 & 14.84 & 1.56 & 6.3 & 6.9 & 1.2 & 7.8 & 0.2 & 0.1 & -- & -- & -- & 0.89(7) & -- & -- & -- & --, --, 18, --\\
J1614$-$2230 & 3.15 & 0.96 & 8.7 & 11.3 & 1.3 & -- & 12.1 & 7.5 & Y & 32.6 & $1.97_{-0.04}^{+0.04}$ & 0.999894(5) & $78_{-8}^{+12}$ & OG & -- & 1, 11, 11, 26\\
J1618$-$39 & 11.99 & -- & 22.8 & 10.2 & 2.7 & -- & -- & -- & -- & -- & -- & -- & -- & -- & $43.3_{-6.1}^{+7.2}$ & --, --, --, --\\
J1622$-$6617 & 23.62 & 6.40 & 1.6 & 1.0 & 2.2 & -- & 0.2 & $3.8 \times 10^{-2}$ & -- & -- & -- & -- & -- & -- & $29.6_{-3.7}^{+4.0}$ & --, --, --, --\\
J1640+2224 & 3.16 & 0.28 & 175.5 & 55.3 & 1.2 & 11.4 & 2.1 & 1.5 & Y & 14 & -- & 0.995(5) & -- & -- & $48.1_{-6.7}^{+8.5}$ & 3, --, 25, --\\
J1643$-$1224 & 4.62 & 1.85 & 147.0 & 25.1 & 0.4 & 7.3 & 7.3 & 41.3 & -- & -- & -- & -- & -- & -- & $22.8_{-2.7}^{+3.0}$ & --, --, --, --\\
J1708$-$3506 & 4.51 & 2.30 & 149.1 & 33.6 & 2.8 & -- & 9.9 & 1.3 & -- & -- & -- & -- & -- & -- & $30.9_{-3.8}^{+4.3}$ & --, --, --, --\\
J1713+0747 & 4.57 & 0.85 & 67.8 & 32.3 & 1.1 & 6.3 & 3.3 & 3.0 & Y & 39 & $1.53_{-0.06}^{+0.08}$ & 0.95(1) & $73_{-9}^{+5}$ & OG & $67.3_{-7.5}^{+15.9}\bigstar$ & 1, 12, 12, 26\\
J1732$-$5049 & 5.31 & 1.42 & 5.3 & 4.0 & 1.4 & -- & 3.7 & 1.9 & Y & 45 & -- & -- & -- & -- & $55.8_{-9.0}^{+12.6}$ & 3, --, --, --\\
J1738+0333 & 5.85 & 2.41 & 0.4 & 0.3 & 1.4 & 6.9 & 4.6 & 2.2 & -- & -- & $1.47_{-0.06}^{+0.07}$ & 0.539(15) & -- & -- & $32.1_{-1.9}^{+2.1}\bigstar$ & --, 13, 13, --\\
J1741+1351 & 3.75 & -- & 16.3 & 11.0 & 0.9 & -- & 21.8 & 18.7 & Y & 1.5 & -- & -- & -- & -- & $73.2_{-15.6}^{+16.8}$ & 1, --, --, --\\
J1744$-$1134 & 4.07 & 0.89 & -- & -- & 0.4 & 21.0 & 4.1 & 23.5 & Y & 16.5 & -- & -- & $85_{-12}^{+3}$ & PSPC & -- & 1, --, --, 26\\
J1745$-$0952 & 19.38 & 9.25 & 4.9 & 2.4 & 1.8 & 23.9 & 0.2 & $7.0 \times 10^{-2}$ & -- & -- & -- & -- & -- & -- & $31.2_{-4.0}^{+4.4}$ & --, --, --, --\\
J1748$-$3009 & 9.68 & -- & 2.9 & 1.3 & 5.1 & -- & -- & -- & -- & -- & -- & -- & -- & -- & $25.4_{-3.1}^{+3.4}$ & --, --, --, --\\
J1751$-$2857 & 3.91 & 1.13 & 110.7 & 32.5 & 1.1 & -- & 7.4 & 6.1 & -- & -- & -- & -- & -- & -- & $39.1_{-5.1}^{+6.0}$ & --, --, --, --\\
J1801$-$3210 & 7.45 & 0.27 & 20.8 & 7.8 & 4.0 & -- & 0.3 & $1.6 \times 10^{-2}$ & -- & -- & -- & -- & -- & -- & $34.2_{-4.5}^{+5.1}$ & --, --, --, --\\
J1802$-$2124 & 12.65 & 7.26 & 0.7 & 3.7 & 2.9 & 4.9 & 1.4 & 0.2 & -- & -- & $1.24_{-0.11}^{+0.11}$ & 0.984(2) & -- & -- & -- & --, 14, 14, --\\
J1804$-$2717 & 9.34 & 4.09 & 11.1 & 7.3 & 0.8 & -- & 2.0 & 3.3 & -- & -- & $1.3_{-0.4}^{+0.4}$ & -- & -- & -- & $50.9_{-13.2}^{+19.9}\bigstar$ & --, 10, --, --\\
J1816+4510 & 3.20 & 4.10 & 0.4 & 0.6 & 2.4 & -- & 49.4 & 8.6 & Y & 25 & -- & -- & -- & -- & $68.6_{-12.8}^{+21.4}$ & 4, --, --, --\\
J1835$-$0114 & 5.12 & 0.70 & 6.7 & 4.7 & 2.7 & -- & 2.1 & 0.3 & -- & -- & -- & -- & -- & -- & $53.5_{-8.4}^{+11.2}$ & --, --, --, --\\
J1841+0130 & 29.77 & 817.00 & 10.5 & 3.5 & 3.6 & -- & 12.2 & 0.9 & -- & -- & -- & -- & -- & -- & $25.4_{-3.2}^{+3.5}$ & --, --, --, --\\
J1844+0115 & 4.19 & 1.07 & 50.6 & 14.2 & 4.0 & -- & 5.8 & 0.4 & -- & -- & -- & -- & -- & -- & $30.6_{-3.9}^{+4.3}$ & --, --, --, --\\
J1850+0124 & 3.56 & 1.09 & 84.9 & 34.0 & 3.4 & -- & 9.5 & 0.8 & -- & -- & -- & -- & -- & -- & $54.3_{-8.3}^{+11.6}$ & --, --, --, --\\
J1853+1303 & 4.09 & 0.87 & 115.7 & 40.8 & 2.1 & 3.4 & 4.9 & 1.1 & -- & -- & $1.4_{-0.7}^{+0.7}$ & -- & -- & -- & $46.7_{-16.8}^{+25.8}\bigstar$ & --, 15, --, --\\
J1857+0943 & 5.36 & 1.78 & 12.3 & 9.2 & 0.9 & 6.1 & 4.5 & 5.5 & -- & -- & $1.57_{-0.11}^{+0.12}$ & 0.9990(7) & -- & -- & $90.0_{-21.7}^{+0.0}\bigstar$ & --, 16, 24, --\\
J1900+0308 & 4.91 & 0.59 & 12.5 & 6.7 & 5.8 & -- & 2.0 & $5.8 \times 10^{-2}$ & -- & -- & -- & -- & -- & -- & $45.9_{-6.6}^{+8.0}$ & --, --, --, --\\
J1902$-$5105 & 1.74 & -- & 2.0 & 1.9 & 1.2 & -- & 68.6 & 49.3 & Y & 5.2 & -- & -- & $42_{-22}^{+17}$ & alTPC & $55.3_{-8.7}^{+11.8}$ & 1, --, --, 26\\
J1903+0327 & 2.15 & 1.88 & 95.2 & 105.6 & 6.4 & 5.6 & 70.6 & 1.7 & -- & -- & $1.667_{-0.021}^{+0.021}$ & 0.9760(15) & -- & -- & -- & --, 17, 17, --\\
J1909$-$3744 & 2.95 & 1.40 & 1.5 & 1.9 & 1.3 & 37.1 & 2.5 & 1.6 & -- & -- & $1.47_{-0.02}^{+0.03}$ & 0.9980(1) & -- & -- & $79.6_{-9.7}^{+10.4}\bigstar$ & --, 18, 24, --\\
J1910+1256 & 4.98 & 0.97 & 58.5 & 21.1 & 2.3 & 7.3 & 2.6 & 0.5 & -- & -- & $1.6_{-0.6}^{+0.6}$ & -- & -- & -- & $44.0_{-12.7}^{+16.7}\bigstar$ & --, 15, --, --\\
J1911$-$1114 & 3.63 & 1.42 & 2.7 & 1.8 & 1.2 & 23.8 & 6.7 & 4.5 & -- & -- & -- & -- & -- & -- & $37.3_{-5.0}^{+5.5}$ & --, --, --, --\\
J1918$-$0642 & 7.65 & 2.57 & 10.9 & 8.4 & 1.2 & 9.2 & 2.1 & 1.4 & -- & -- & -- & -- & -- & -- & $82.0_{-21.2}^{+8.0}$ & --, --, --, --\\
J1935+1726 & 4.20 & -- & 90.8 & 32.0 & 3.2 & -- & -- & -- & -- & -- & -- & -- & -- & -- & $46.5_{-6.5}^{+8.1}$ & --, --, --, --\\
J1939+2134 & 1.56 & 10.51 & -- & -- & 5.0 & 0.8 & 1097.6 & 43.9 & Y & 1.3 & -- & -- & $82_{-32}^{+8}$ & alTPC & -- & 1, --, --, 26\\
J1949+3106 & 13.14 & 9.39 & 1.9 & 7.3 & 6.5 & 5.9 & 1.5 & $3.5 \times 10^{-2}$ & -- & -- & $1.47_{-0.31}^{+0.43}$ & 0.985(6) & -- & -- & -- & --, 19, 19, --\\
J1955+2908 & 6.13 & 2.97 & 117.3 & 31.4 & 4.6 & 4.2 & 4.9 & 0.2 & -- & -- & -- & -- & -- & -- & $35.6_{-4.5}^{+5.2}$ & --, --, --, --\\
J1959+2048 & 1.61 & 1.69 & 0.4 & 0.1 & 2.5 & 30.4 & 75.0 & 12.1 & Y & 16.5 & -- & -- & $83_{-31}^{+7}$ & alTPC & -- & 1, --, --, 26\\
J2017+0603 & 2.90 & 0.83 & 2.2 & 2.2 & 1.6 & -- & 13.5 & 5.5 & Y & 75.5 & -- & -- & $68_{-5}^{+8}$ & OG & $62.4_{-10.9}^{+21.0}$ & 1, --, --, 26\\
J2019+2425 & 3.93 & 0.70 & 76.5 & 38.8 & 1.5 & 22.6 & 4.6\dag & 2.1\dag & -- & -- & $1.205_{-0.305}^{+0.305}$ & -- & -- & -- & $62.3_{-15.1}^{+27.7}\bigstar$ & --, 20, --, --\\
J2033+1734 & 5.95 & 1.11 & 56.3 & 20.2 & 2.0 & 12.5 & 1.2 & 0.3 & -- & -- & -- & -- & -- & -- & $41.7_{-5.7}^{+6.7}$ & --, --, --, --\\
J2043+1711 & 2.38 & 0.52 & 1.5 & 1.6 & 1.8 & 13.0 & 10.3 & 3.3 & Y & 79 & $1.85_{-0.15}^{+0.15}$ & 0.990(8) & $78_{-7}^{+2}$ & TPC & $81.5_{-16.7}^{+8.5}\bigstar$ & 1, 21, 21, 21\\
J2124$-$3358 & 4.93 & 2.06 & -- & -- & 0.3 & 52.3 & 3.5 & 39.3 & Y & 10.8 & -- & -- & $20_{-8}^{+5}$ & PSPC & -- & 1, --, --, 26\\
J2129$-$5721 & 3.73 & 2.09 & 6.6 & 3.5 & 0.4 & 13.3 & 15.4 & 96.4 & -- & -- & -- & -- & -- & -- & $37.6_{-5.0}^{+5.7}$ & --, --, --, --\\
J2214+3000 & 3.12 & 1.40 & 0.4 & 0.1 & 1.5 & -- & 18.2 & 7.7 & Y & 48.4 & -- & -- & $14_{-8}^{+46}$ & alTPC & -- & 1, --, --, 26\\
J2229+2643 & 2.98 & 0.15 & 93.0 & 18.9 & 1.4 & 17.0 & 2.2\dag & 1.0\dag & -- & -- & -- & -- & -- & -- & $24.9_{-3.0}^{+3.3}$ & --, --, --, --\\
J2302+4442 & 5.19 & 1.33 & 125.9 & 51.4 & 1.2 & -- & 3.8 & 2.7 & Y & 162.6 & -- & -- & $46_{-7}^{+7}$ & TPC & $64.2_{-11.2}^{+25.8}$ & 1, --, --, 26\\
J2317+1439 & 3.45 & 0.24 & 2.5 & 2.3 & 0.8 & 7.6 & 1.9 & 2.8 & -- & -- & -- & -- & -- & -- & $59.2_{-9.9}^{+15.2}$ & --, --, --, --\\

\end{longtable}
\end{tiny}
\end{center}
\end{landscape}
\twocolumn

\end{document}